\documentclass[aps,prb,twocolumn,nopacs,superscriptaddress]{revtex4}
\usepackage[breaklinks=true,colorlinks,citecolor=blue,linkcolor=blue,urlcolor=blue]{hyperref}
\usepackage{epsfig,mathrsfs,color,latexsym,subfigure,marginnote}

\usepackage{graphicx}
\usepackage{verbatim}
\usepackage{relsize}
\usepackage{mathrsfs}
\usepackage{color}
\usepackage{array}
\usepackage{amsmath,amsfonts,amssymb}
\usepackage{graphicx}
\renewcommand{\BibitemShut}[1]{}

\newcommand{\be}{\begin{equation}}
\newcommand{\ee}{\end{equation}}
\newcommand{\ba}{\begin{eqnarray}}
\newcommand{\ea}{\end{eqnarray}}

\begin{document}
\title{Starfruit-like nodal semimetal to Dirac and Weyl semimetal state in CaAuAs}

\author{Bahadur Singh\footnote{These authors contributed equally to this work.}$^{\dag}$}
\affiliation{SZU-NUS Collaborative Center and International Collaborative Laboratory of 2D Materials for Optoelectronic Science $\&$ Technology, Engineering Technology Research Center for 2D Materials Information Functional Devices and Systems of Guangdong Province, College of Optoelectronic Engineering, Shenzhen University, ShenZhen 518060, China}
\affiliation{Department of Physics, Northeastern University, Boston, Massachusetts 02115, USA}

\author{Sougata Mardanya$^*$}
\affiliation{Department of Physics, Indian Institute of Technology - Kanpur, Kanpur 208016, India}

\author{Chenliang Su}
\affiliation{SZU-NUS Collaborative Center and International Collaborative Laboratory of 2D Materials for Optoelectronic Science $\&$ Technology, Engineering Technology Research Center for 2D Materials Information Functional Devices and Systems of Guangdong Province, College of Optoelectronic Engineering, Shenzhen University, ShenZhen 518060, China}

\author{Hsin Lin$^{\dag}$}
\affiliation{Institute of Physics, Academia Sinica, Taipei 11529, Taiwan}

\author{Amit Agarwal\footnote{Corresponding authors' emails: 
bahadursingh24@gmail.com, nilnish@gmail.com, amitag@iitk.ac.in}}
\affiliation{Department of Physics, Indian Institute of Technology - Kanpur, Kanpur 208016, India}

\author{Arun Bansil}
\affiliation{Department of Physics, Northeastern University, Boston, Massachusetts 02115, USA}

\begin{abstract}
Band-crossings occurring on a mirror plane are compelled to form a nodal loop in the momentum space without spin-orbit coupling (SOC). In the presence of other equivalent mirror planes, multiple such nodal loops can combine to form interesting nodal-link structures. Here, based on first-principles calculations and an effective $\mathbf{k.p}$ model analysis, we show that CaAuAs hosts a unique starfruit-like nodal-link structure in the bulk electronic dispersion in the absence of SOC. This nodal-link is comprised of three nodal loops, which cross each other at the time-reversal-invariant momentum point $A$. When the SOC is turned on, the nodal loops are gapped out, resulting in a stable Dirac semimetal state with a pair of Dirac points along the $\mathrm{\Gamma-A}$ direction in the Brillouin zone. The Dirac points are protected by the combination of time reversal, inversion, and $C_3$ rotation symmetries. We show how a systematic elimination of the symmetry constraints yields a Weyl semimetal and eventually a topological insulator state.  

\end{abstract}
\maketitle

\section{Introduction} 

Topological phases of quantum matter are currently at the forefront of condensed matter and materials sciences research\cite{Bansil2016,Hasan2010,Ashvin2018, kong11}. The initial focus on insulating phases has shifted in the last few years to topological semimetals in which the bulk electronic spectrum hosts symmetry-protected gapless crossings near the Fermi level between the valence and conduction bands. Electronic states near the band-crossings mimic the Dirac and Weyl fermions familiar in the standard model of relativistic high-energy physics. The related Dirac and Weyl semimetals possess discrete four- and two-fold degenerate protected gapless points in their bulk spectra in which low energy excitations are Dirac\cite{young12,young15,weider16,liu14a,liu14b,neupane14,borisenko14,wang12,wang13,zhang17,chen17} and Weyl fermions\cite{xu15a,huang15,lv15,xu15b,xu11,fang12,weng15a,bahadur12,PhysRevLett.119.026404,chiu14,gao16}. Under certain symmorphic symmetry operations such as mirror planes, the band-crossings in many cases persist along one-dimensional (1D) loops, and give rise to single-nodal-line semimetals (NLSM), or multiple NLSMs, nodal-chain or topological Hopf-link semimetals \cite{fang16,burkov11b,phillips14,fang15,bian16a,neupane16,schoop16,xie15,kim15,mullen15,weng15b,yu15,chen15,yamakage15,li16,jin17,bian16b,bahadur17,bahadur18}. In contrast to Dirac and Weyl semimetals, energy dispersion in the NLSMs is quadratic in momentum along the tangential direction to the nodal line and linear in momentum along the other two perpendicular directions. Low-energy excitations in the NLSMs with highly anisotropic energy dispersions do not have any high-energy counterparts. Nontrivial bulk band topologies of topological semimetals are also associated with 2D surface states with open Fermi surfaces in the Weyl semimetals and flat drumhead-like Fermi surfaces in the NLSMs. The unique bulk and surface states of topological semimetals provide a new platform for investigating various intriguing high-energy and relativistic physics phenomena in table-top experiments and provide an exciting basis for developing next-generation technological applications\cite{Ashvin2018,Hasan2010,Bansil2016, kong11}.

Among the topological semimetals, the NLSMs are perhaps more interesting because of their high density of states (DOS) at the Fermi level which could drive exotic correlation physics in these materials\cite{bahadur18,nodal_inter2016}. The nontrivial band structure of NLSMs leads to distinct magnetic, optical, and transport properties compared to that of Weyl and Dirac semimetals. NLSMs have been theoretically proposed in various families of compounds and experimentally verified recently in PbTaSe$_2$, CaAgAs, and ZrSi(S,Te) \cite{bian16a,daichi18,neupane16,schoop16,hu16,yamakage15,lodge17}. A focus of research has been the stability of nodal-line crossings with respect to the strength of the SOC, whose presence usually unlocks nodal loops to yield various topological states such as the Dirac and Weyl semimetals and the fully gapped insulators. Moreover, since nodal lines are enforced by mirror symmetries, they can form interesting linked nodal-line connections in the presence of multiple mirror planes\cite{kobayashi17}.  

In this paper, we discuss the topological NLSM state and its transition to the Dirac and Weyl semimetal states of ternary hexagonal CaAuAs using first-principles band structure calculations and a $\mathbf{k.p}$ model Hamiltonian analysis. We show that CaAuAs realizes a unique nodal link semimetal state, which is composed of three nodal rings in the absence of SOC. The nodal lines are located on the vertical mirror symmetry planes of the $D^4_{6h}$ symmetry group and intersect at the $A$ point on the momentum space Brillouin zone (BZ), forming a starfruit-like nodal connection. A similar nodal-link semimetal state has been reported recently in $\mathrm{YH_3}$\cite{kobayashi17} and LiAuSe\cite{chen17b}. Inclusion of the SOC gaps out the nodal lines and transitions the system into a stable, symmetry-protected Dirac semimetal state with a pair of Dirac points, which are located along the $C_3$ rotation axis. We also characterize the topological state of CaAuAs through surface-state calculations. It is well known that Dirac semimetals can undergo tunable transitions to a variety of gapped as well as semimetal phases via external magnetic fields. Accordingly, we have investigated topological transitions in CaAuAs by breaking $C_3$ rotational and time-reversal symmetries.

The remainder of this paper is organized as follows. In Sec.~\ref{CS_CD}, we describe computational details and discuss the crystal structure and symmetry properties of CaAuAs. Sec.~\ref{NLSM} discusses the starfruit like NLSM state. In Sec.~\ref{DSM}, we present an analysis of the SOC-induced Dirac semimetal state. The topological surface electronic structure is also explored. In Sec.~\ref{KP}, we present an effective $\mathbf{k\cdot p}$ Hamiltonian. Effects of selective symmetry-breaking and the emergence of topological insulator and Weyl semimetal states are discussed in Sec.~\ref{broken}. Finally, we summarize our findings in Sec.~\ref{Conc}.

\section{Methodology and Crystal structure} \label{CS_CD}

Electronic structure calculations were performed within the framework of the density functional theory (DFT) with the projector augmented wave (PAW) pseudopotentials\cite{paw94}, as implemented in the Vienna \textit{ab initio} simulation (VASP) package \cite{kresse96, kresse99}. The generalized gradient approximation with the Perdew-Burke-Ernzerhof parameterization was used to include exchange-correlation effects\cite{gga96}. Bulk calculations used a plane wave energy cut-off of 500 eV and a $20\times20\times20$ $\Gamma$-centered $k-$mesh to sample the BZ\cite{mpgrid76}. Total energies were converged to $10^{-7}$ eV. The SOC was used self-consistently to incorporate relativistic effects. Experimental lattice parameters of CaAuAs ($a=b=4.388 \AA$ and $c=7.925 \AA$) were used. Topological properties were calculated by employing a tight-binding model obtained by using the WannierTools suite of codes\cite{wannier90,wanniertools}. VESTA\cite{vesta} software package was used for crystal structure visualization.

\begin{figure}[th!]
\includegraphics[width=0.90 \linewidth]{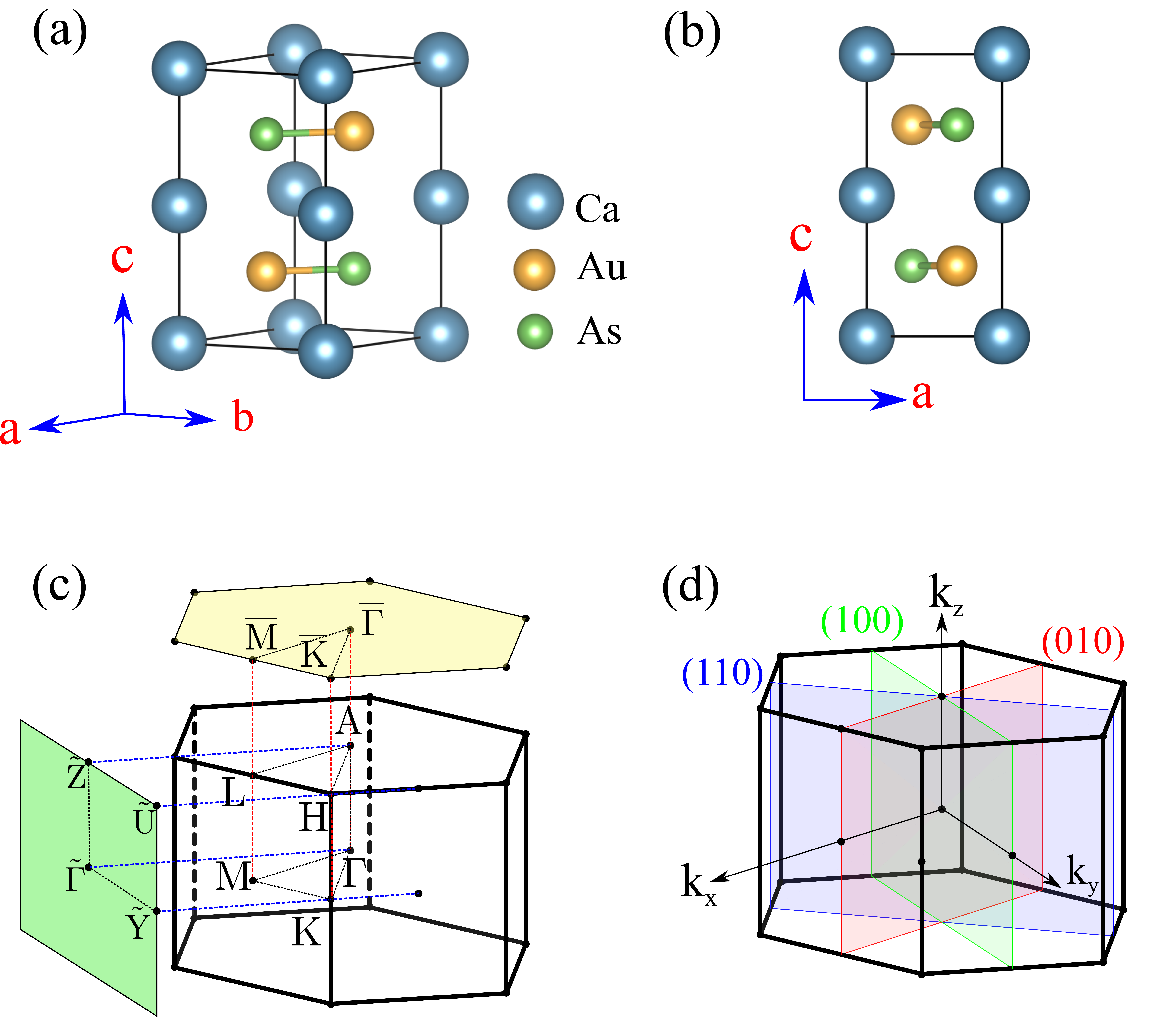}
\caption{(a) Side and (b) (010) view of the crystal structure of CaAuAs. (c) Bulk BZ and its projection onto the (001) and (010) surface planes. The relevant bulk and surface high-symmetry points are marked. (d) Illustration of three mirror-reflection planes $m_{100}$ (red), $m_{010}$ (green) and $m_{110}$ (blue) in the bulk BZ.}\label{fig:CS}
\end{figure}

CaAuAs crystallizes in the hexagonal crystal lattice with space group $D^4_{6h}$ (No. 194). The primitive unit cell and its (010) view are shown in Figs. \ref{fig:CS}(a)-(b). The unit cell contains six atoms (two Ca, two Au, and two As), which occupy Wyckoff positions 2a, 2d, and 2d. The crystal structure can be viewed as a shared honeycomb lattice of Au and As atoms that are stacked along the hexagonal $z$-axis. The hexagonal Ca layers are inserted in this stacking sequence while maintaining the bulk inversion symmetry. The bulk BZ and the associated surface-projected BZs are shown in Fig. \ref{fig:CS}(c). The crystal has three equivalent mirror-reflection planes $m_{110}$, $m_{100}$, and $m_{010}$, which are illustrated in Fig. \ref{fig:CS}(d).

\section{A starfruit-like nodal-line semimetal} \label{NLSM}

\begin{figure*}[t]
\begin{center}
\includegraphics[width=0.90 \textwidth]{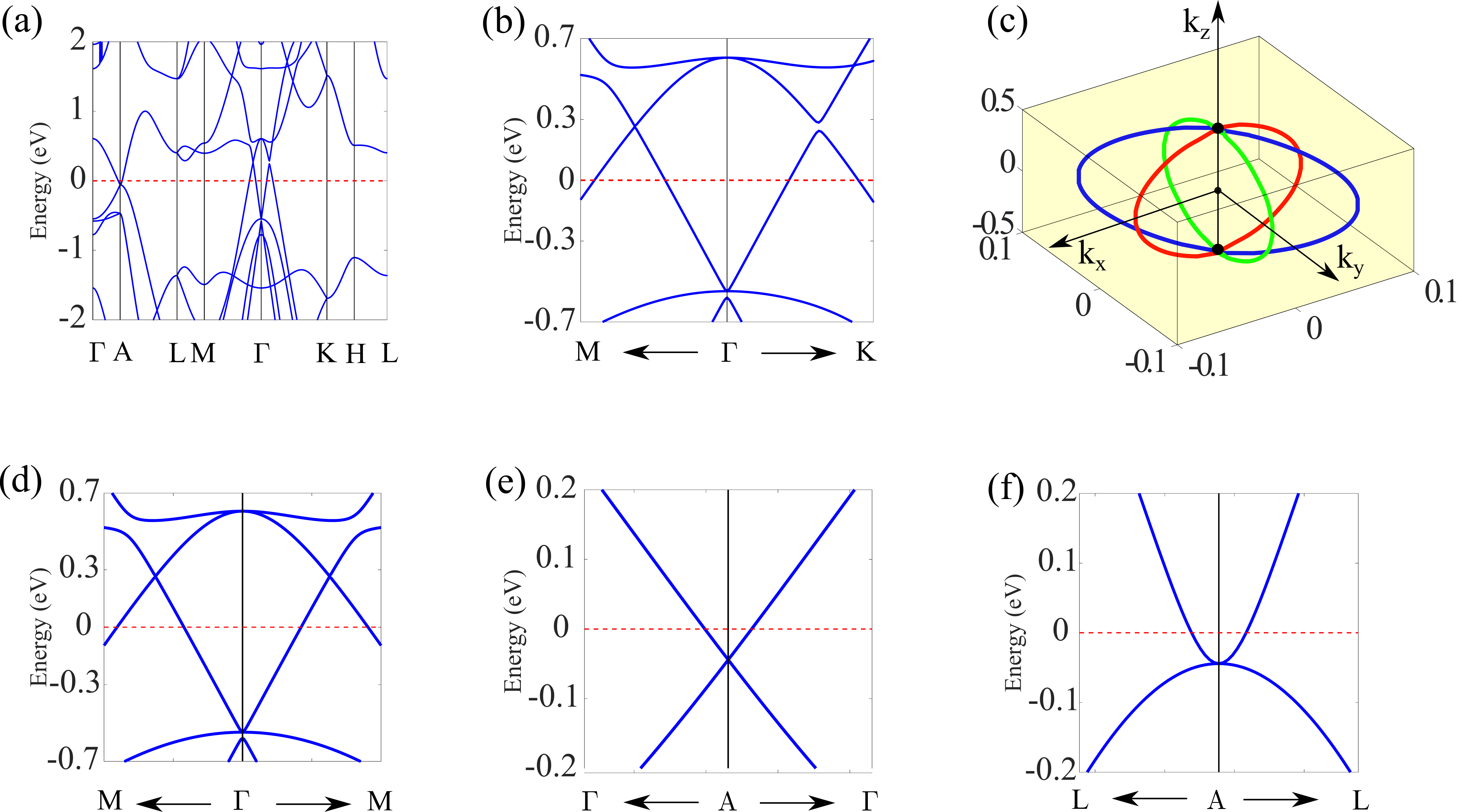}
\caption{(a) Electronic band structure of CaAuAs along the high-symmetry directions in the BZ without SOC. (b),(d) Closeups of the band-structure along the $M-\Gamma-K$ path. A linearly dispersing gapless point, which lies on the $m_{100}$ mirror plane, is evident along the $\Gamma-M$ direction around 0.26 eV. (c) Configuration of nodal lines in the full BZ. Nodal lines located on the three equivalent mirror planes cross at the $\pm A$ point. (e)-(f) Energy dispersion in the vicinity of the $A$ point: It is linear along the $\Gamma-A$ ($k_z$) direction and non-linear in the $L-A$ ($k_x$, $k_y$) directions.}\label{fig:nodal}
\end{center}
\end{figure*}

Electronic structure of CaAuAs without SOC is shown in Fig. \ref{fig:nodal}. Band-crossings in inversion symmetric solids usually occur on high-symmetry lines or planes. The stability of these crossings against `band repulsion' arises from the fact that the bands that are crossing belong to different irreducible representations of the space group. In particular, on a mirror symmetry plane, if the two bands involved have opposite mirror eigenvalues, then they are constrained by this symmetry to form a stable nodal loop. These nodal lines, which appear on equivalent mirror planes, can easily get linked at high-symmetry points or intersection of mirror planes to form chain-like structures.

Figure ~\ref{fig:nodal}(a) shows the band structure of CaAuAs along the selected directions in the bulk BZ. A symmetry analysis shows that, near the Fermi energy, the valence and conduction bands at $\Gamma$ belong to $E_{2g}$ and $E_{1u}$ representations of the $D_{6h}$ point group, respectively. These bands have opposite mirror eigenvalues and they cross along the $\Gamma-M$ direction as shown in the closeups of Figs.~\ref{fig:nodal}(b) and (d). Note that the $\Gamma-M$ direction is an irreducible line associated with the three equivalent mirror planes highlighted in Fig. \ref{fig:CS}(a). A careful analysis reveals that this band crossing persists along the mirror plane to form a closed loop. Further exploration of the 3D band structure shows the presence of similar nodal lines on the $m_{010}$ and $m_{110}$ mirror planes, see Fig.~\ref{fig:nodal}(c). These equivalent nodal loops are linked at the high-symmetry point $A = ({0,0,\pm \pi})$ and form a starfruit-like nodal structure [see Fig.~\ref{fig:nodal}(c)]. \footnote{A similar nodal structure has been reported recently in the rare-earth-trihalide YH$_3$\cite{kobayashi17}. The conduction and valence bands in YH$_3$ at the $\Gamma$ point belong to A$_{2u}$ and A$_{2g}$ symmetries, respectively, with the crossing-point of nodal loops lying on the $\Gamma-A$ line.} CaAuAs also exhibits a $C_3$ rotation-symmetry-protected Dirac point at the $A$-point with a unique eight-fold degeneracy without the SOC; the band dispersion around this point is anisotropic, being linear along the $\mathrm{\Gamma-A}$ direction and non-linear along the in-plane $\mathrm{A-L}$ direction, see Figs.~\ref{fig:nodal}(e) and (f).

\section{Spin-orbit coupling induced Dirac semimetal} \label{DSM}

\begin{figure*}[t]
\begin{center}
\includegraphics[width=0.90 \textwidth]{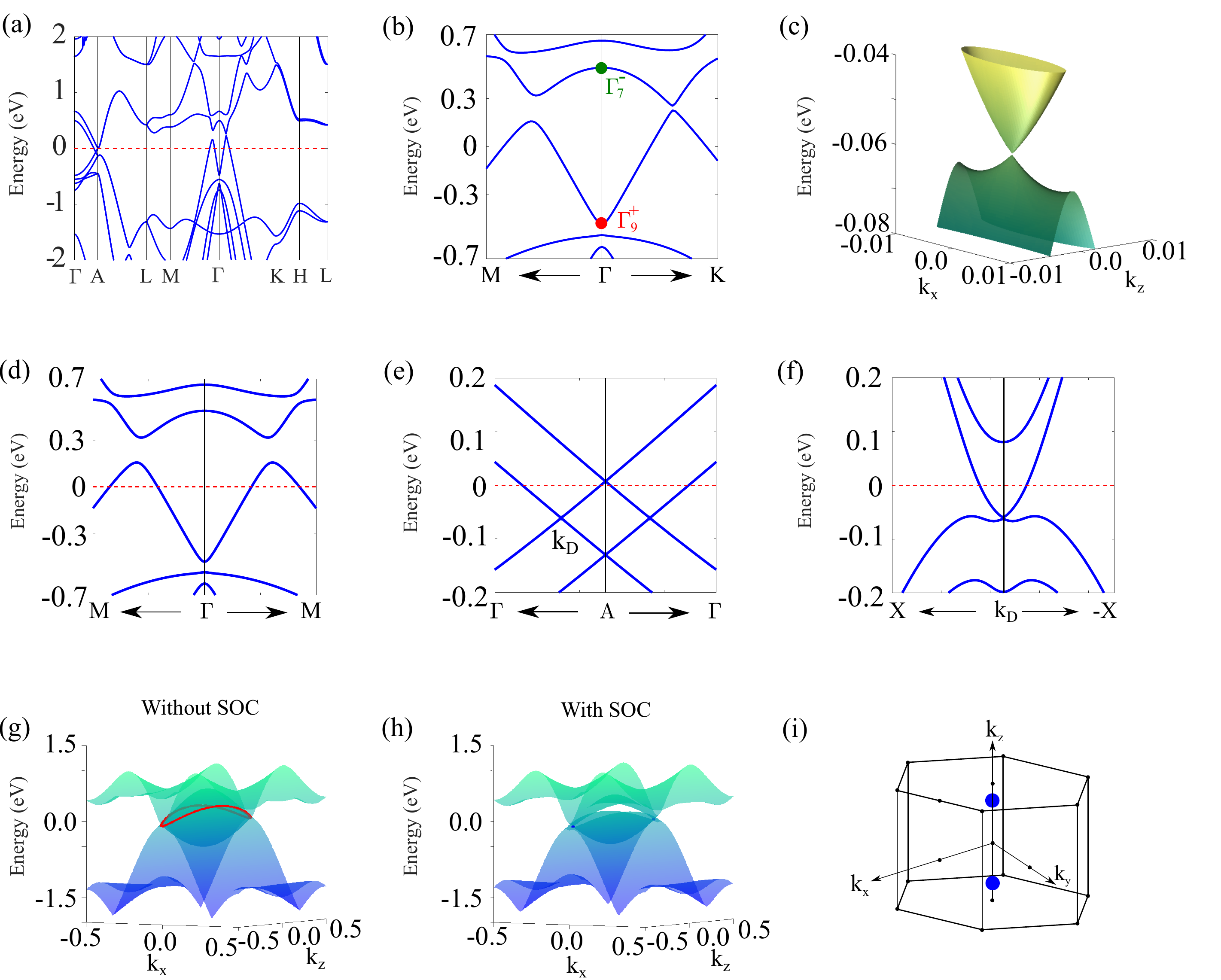}
\caption{(a) Electronic band structure of CaAuAs with SOC. (b),(d) Closeups of the band structure along the high-symmetry directions. A clear band gap emerges at the band-crossings without the SOC on the mirror-symmetric $\Gamma-M$ line. (e) Energy dispersion along the $\Gamma-A$ line displaying a pair of Dirac points at $\pm {\bf k}_D$; dispersion along an in-plane direction is shown in (f). (c) A 3D rendition of the Dirac cone in the $E-k_x-k_z$ space. A linear dispersion along the three principal directions confirms that this is a type-I Dirac cone. (g) 3D dispersion of a single nodal line on the $m_{010}$ plane without the SOC. When the SOC is included, the nodal line is gapped, leaving a pair of unavoidable crossings on the $C_3$ rotation axis as seen in (h). (i) Blue markers show the location of the Dirac points in the BZ. }\label{fig:DSM}
\end{center}
\end{figure*}

In the absence of non-symmorphic symmetries, the SOC can gap nodal lines and drive changes in band topologies. In order to delineate effects of the SOC, we present the band structure of CaAuAs including the SOC in Fig.~\ref{fig:DSM}. The nodal-line crossings are now gapped, generating a clear band gap at the band crossing points as shown in Figs.~\ref{fig:DSM}(b) and (d). However, in contrast to ZrPtGe or ZrSiS \cite{bahadur18,singha17,hu16,lodge17}, here the band crossing is retained only along the $C_3$ rotation axis ($\mathrm{\Gamma-A}$ direction) at $\pm {\bf k}_D$ [see Fig.~\ref{fig:DSM}(e)]. Owing to the presence of both inversion and time-reversal symmetries, the $\pm {\bf k}_D$ points are fourfold degenerate with a linear dispersion along all three directions [Fig. \ref{fig:DSM}(e) and (f)]. A careful symmetry analysis shows that the crossing bands belong to different irreducible representations ($\Gamma_7^{-}$ and $\Gamma_9^{+}$) with opposite parities and $C_3$ rotation eigenvalues. These nodal points are thus unavoidable and protected by the $C_3$ rotation axis. Keeping in mind that the bandgap is typically underestimated in the GGA, we have also computed the band structure using the hybrid exchange-correlation functional to confirm the presence of the aforementioned Dirac semimetal state with a pair of Dirac cones (results not shown for brevity). These results show that CaAuAs realizes a robust, nearly ideal type-I Dirac semimetal in which the Dirac points lie close to the Fermi level, much like the case of Na$_3$Bi. The transition from a linked NLSM to a Dirac semimetal with SOC is illustrated in Figs.~\ref{fig:DSM}(g) and (h).

\begin{figure*}[t]
\begin{center}
\includegraphics[width=0.9 \linewidth]{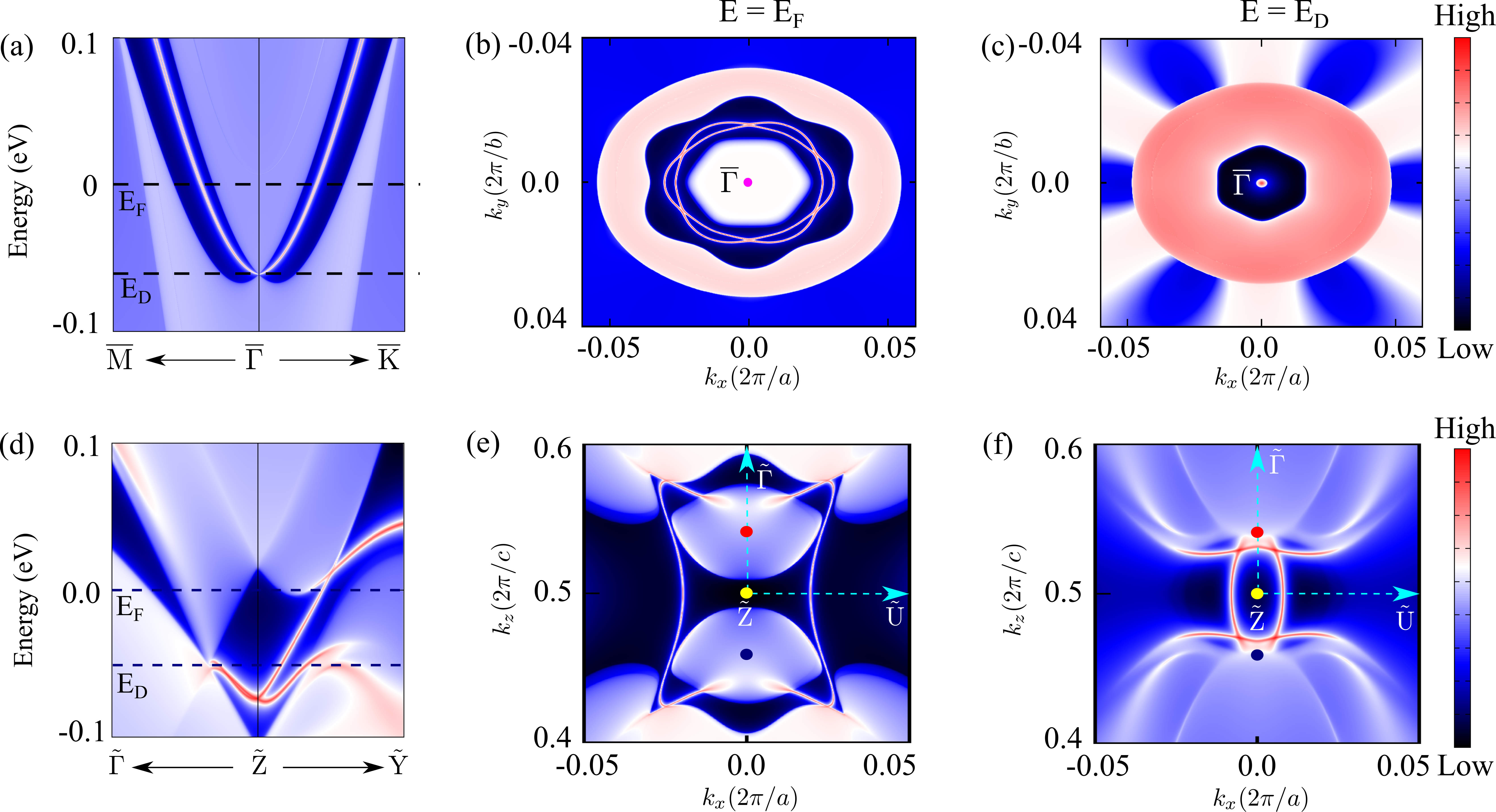}
\caption{(a) Band structure of (001) surface of CaAuAs including SOC. The shaded blue region shows the projected bulk bands and sharp lines identify the surface states. Constant energy contours at (b) $E=E_f$ and (c) $E=E_D$; these energies are marked with dashed lines in (a). (d-f) Same as (a-c) but for (010) surface. The surface states in (c) emanate at the Dirac point along the $\tilde{\Gamma}-\tilde{Z}$ direction. The double-Fermi-arc states connecting the pair of Dirac points is evident from (e) and (f).}\label{fig:SS}
\end{center}
\end{figure*}

The nontrivial bulk band topology is accompanied by the existence of topological surface states. In order to showcase these states and their connection to the bulk bands, we present the (001) and (010) surface band structure of CaAuAs in Fig.~\ref{fig:SS}. On the (001) surface, a pair of Dirac points located on the $\Gamma-A$ bulk direction projects onto the surface $\overline{\Gamma}$ point, which is resolved in Fig. \ref{fig:SS}(a). In addition to these projected bulk Dirac cones, we find topological states that connect the valence and conduction bands as in the case of a topological insulator. These states result from the inverted bulk band structure and may be considered a precursor of the metallic surface states in a topological insulator. Figures \ref{fig:SS}(b) and \ref{fig:SS}(c) show the constant energy cuts corresponding at the Fermi energy ($E_f$) and the energy of the Dirac point ($E_D$), respectively. The projected Dirac points are seen at the $\overline\Gamma$ in Fig. \ref{fig:SS}(c).

Figure \ref{fig:SS}(d) shows the band structure of the (010) surface. The pair of Dirac points on this surface projects onto the $\tilde{\Gamma}-\tilde{Z}$ surface direction [see Fig. \ref{fig:CS}(c)]. We observe surface states that emerge from the Dirac node, suggesting the existence of double Fermi arc states over this surface. Note that a Dirac point is a stable merger of two Weyl points of opposite chiral charge in the presence of a crystalline symmetry. Therefore, double Fermi arcs may connect a pair of Dirac nodes, forming a closed surface loop mediated by the Dirac nodes. Figs. \ref{fig:SS}(e) and \ref{fig:SS}(f) show constant energy contours at $E=E_f$ and $E=E_D$, respectively. We can clearly see a pair of Fermi arcs terminated on the Dirac nodes in \ref{fig:SS}(f). They emerge from one Dirac node on the $\tilde{\Gamma}-\tilde{Z}$ direction and terminate on the other. The evolution of these Fermi arc states at a higher energy is shown in Fig. \ref{fig:SS}(e).

\section{Effective Hamiltonian} \label{KP}
We now discuss a low-energy $\mathbf{k.p}$ effective model Hamiltonian, which is derived using the theory of invariants in a manner similar to Na$_3$Bi \cite{wang12} and LiAuSe\cite{chen17b}. As discussed above, the irreducible representations of the two bands crossing at the $\Gamma$ point are $\Gamma_7^{-}$ and $\Gamma_9^{+}$ with the major contribution from As-p and Au-s orbitals, respectively. Here, superscript $(+)$ and $(-)$ represents parity of the state. In the presence of SOC, a $4\times4$ effective Hamiltonian in the basis $|s^+_{\frac{1}{2}},\frac{1}{2}\rangle$, $|p^-_{\frac{3}{2}},\frac{3}{2}\rangle$, $|s^+_{\frac{1}{2}},-\frac{1}{2}\rangle$, $|p^-_{\frac{3}{2}},-\frac{3}{2}\rangle$ with the constraints of $D^4_{6h}$ symmetries is,

\begin{equation} \label{eq2}
H_\Gamma({\bf k})=\epsilon({\bf k})+\left(\begin{matrix}
M({\bf k}) & A({\bf k}){\bf k}_+ & 0 & -B^*({\bf k})\\
A({\bf k}){\bf k}_- & -M({\bf k}) & B^*({\bf k}) & 0\\
0 & B({\bf k}) & M({\bf k}) & A({\bf k}){\bf k}_-\\
-B({\bf k}) & 0 & A({\bf k}){\bf k}_+ & -M({\bf k})
\end{matrix}\right)~. 
\end{equation}
Here, ${\bf k}_\pm=k_x\pm i k_y$, $\epsilon({\bf k})= c_0 + c_1 k_z^2+c_2(k_x^2+k_y^2)$, $A({\bf k}) = A_0 + A_1 k_z^2+A_2(k_x^2+k_y^2)$, $B({\bf k}) = B_3k_zk_+^2$, and $M({\bf k}) = -M_0 + M_1 k_z^2+M_2(k_x^2+k_y^2)$  with $M_0$, $M_1$, and $M_2 > 0$ to ensure a band inversion. 
The associated energy dispersion of this Hamiltonian is,
\begin{equation} \label{eq3}
E({\bf k})=\epsilon({\bf k})\pm\sqrt{M({\bf k})^2+A({\bf k})^2k_+k_-+|B({\bf k})|^2}~, 
\end{equation}
with a pair of gapless Dirac points located on the $\Gamma-$A line at $\pm {\bf k}_D=(0,0,\pm\sqrt{{M_0}/{M_1}})$. By neglecting the higher order terms in $B(k)=B_3k_zk^2_+$ and $A({\bf k})$, one can obtain a linearized massless Dirac Hamiltonian around each gapless point. By fitting the energy dispersion of Eq. \ref{eq3} with our first-principles results of CaAuAs, we obtain the fitting parameters: $c_0 = -0.05$ eV , $c_1= -0.161$ eV \AA$^2$, $c_2 = 13.127$ eV \AA$^2$, $M_0 = 0.348$ eV, $M_1 = 2.722$ eV \AA$^2$, $M_2 = 13.452$ eV \AA$^2$, $A_0 = 1.182$ eV, $A_1 = 4.236$ eV \AA$^2$, $A_2 = -15.598$ eV \AA$^2$ and $B_3 = 0.0004$ eV \AA$^4$.   

\section{Broken-symmetry-induced topological states in $\mathbf{CaAuAs}$}
\label{broken}
Recall that a Dirac semimetal can be thought of as providing a bridge to a variety of topological phases, which can be obtained when the underlying symmetries are broken. We have shown above that the Dirac points in CaAuAs are protected by the $C_3$ rotation symmetry in an inversion- and time-reversal symmetric environment. We now discuss the novel physics that could be realized by breaking symmetries in CaAuAs.

\subsection{Topological insulator}
Breaking the $C_3$ rotation symmetry in CaAuAs introduces an additional linear leading order term in $B({\bf k})=B_1k_z$ in Eq. \ref{eq2}, which ensures gap opening at the Dirac points. To confirm this effect independently, we applied a compressive in-plane strain by varying the angle between the in-plane hexagonal lattice vectors from $120^{\circ}$ to $116^{\circ}$ in our {\it ab-initio} calculations. This breaks the ${C_3}$ symmetry and indeed leads to a gap at the Dirac points as shown in Fig.~\ref{fig:WS}(a). Since Dirac semimetal state in CaAuAs arises through a bulk band inversion, the $C_3$ symmetry broken state could realize a topological insulator with Dirac cone surface states. In Fig. \ref{fig:WS}(b) we show the energy dispersion of (001) surface, which confirms that this is in fact the case with the presence of a single Dirac cone that spans the bulk energy spectrum.

\subsection{Topological Weyl semimetal}

\begin{figure}[h!]
\begin{center}
\includegraphics[width=0.99 \linewidth]{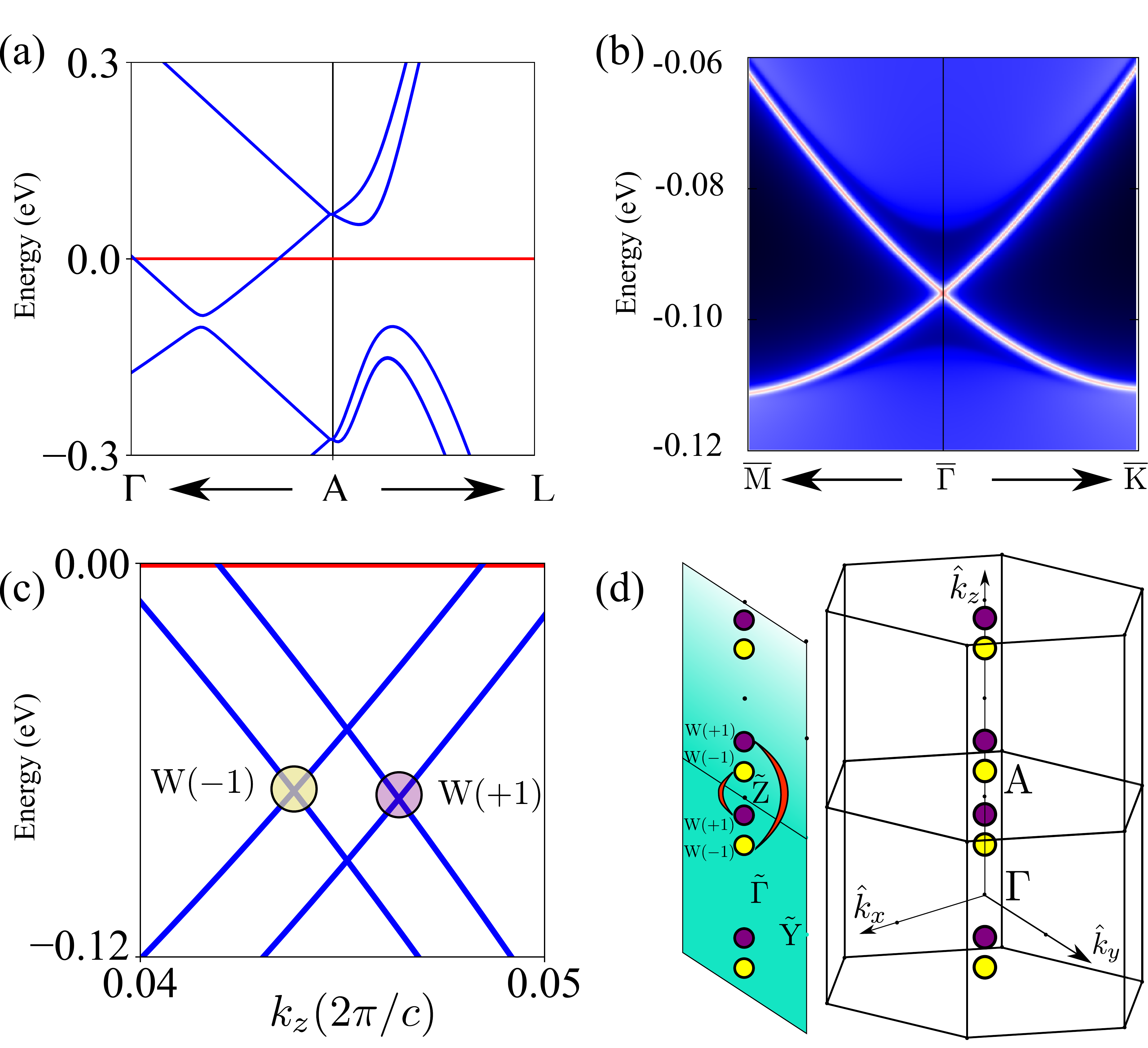}
\caption{(a) Bulk band structure of CaAuAs when the $C_3$ rotational symmetry is broken. A small band gap appears at the Dirac point along the $\Gamma-A$ direction. The associated surface energy spectrum is shown in (b). The topological surface Dirac cone is seen within the bulk band gap. (c) Bulk band structure of CaAuAs in the reduced $k_z$ direction with the Zeeman field [$h=0.02$ eV in Eq.~\eqref{WSM_Ham}].  A pair of Weyl points with opposite chiral charges is shown. (d) A schematic of the location of Weyl points and the Fermi-arc surface states in the bulk and surface BZs.}\label{fig:WS}
\end{center}
\end{figure}

Topological Weyl semimetal is the most robust topological state of quantum matter in the sense that it requires only the transitional invariance of the crystal. In CaAuAs, the Weyl semimetal can be achieved by breaking either the inversion-symmetry or the time-reversal symmetry. Here, we illustrate this evolution of the Dirac semimetal through breaking the time-reversal symmetry by introducing a Zeeman field along the $z$ direction in the low-energy Hamiltonian. The model Hamiltonian of Eq.~\eqref{eq2} with the Zeeman field leads to 
\begin{equation} \label{WSM_Ham}
H_{\rm WSM}(\mathbf{k}) = H_{\Gamma}(\mathbf{k}) + h\sigma_z \tau_z. 
\end{equation}
Here, the second term describes the effective Zeeman field with strength $h$. As a result, each Dirac node at  
$k_{\rm D} = (0,0,\pm\sqrt{{M_0}/{M_1}})$ with four-fold band degeneracy is found to split into a pair of Weyl points as shown in Fig.~\ref{fig:WS}(c).
These Weyl points appear on the $\Gamma-A$ direction at $\pm k_{\rm W}^{\pm} = (0,0,\sqrt{{(M_0 \pm h)}/{M_1}})$.

In order to evaluate the chirality associated with each of these four Weyl nodes, we use our effective Hamiltonian $H_{\rm WSM}$ in the vicinity of the Weyl points. Neglecting the fourth order off-diagonal terms in Eq.~\eqref{WSM_Ham}, the upper diagonal block can be expressed as  
\begin{equation} \label{eq5}
H^{(2 \times 2)}_{\rm WSM}(\mathbf{k})  =  f_0(\mathbf{k})\mathbb{I} + f_1(\mathbf{k})\sigma_1 +f_2(\mathbf{k})\sigma_2 + f_3(\mathbf{k})\sigma_3~.
\end{equation}
Here, $\mathbb{I}$ is the $2 \times 2$ identity matrix, $\sigma_{i}$ denote the three Pauli matrices, $f_0(\mathbf{k}) = \epsilon(\mathbf{k})$, $f_1(\mathbf{k}) = A(\mathbf{k}) k_x$, $f_2(\mathbf{k}) = -A(\mathbf{k}) k_y$ and $f_3(\mathbf{k}) = M(\mathbf{k}) + h $. 
This Hamiltonian results in a pair of gapless crossings at  $\pm \mathbf{k}^{-}_{\rm W}= \pm \left(0,0,\sqrt{\frac{M_0-h}{M_1}}\right)$. By doing a wave vector expansion in the vicinity of each Weyl node i.e,  $\mathbf{k} \to \pm \mathbf{k}^{-}_{\rm W} + \delta \mathbf{k} $, we obtain
\begin{equation} \label{eq4}
H^{(2 \times 2)}_{\rm WSM}(\mathbf{k}) \approx f_0(\pm \mathbf{k}^{-}_{\rm W})\mathbb{I} + \mathbf{v}_0. \delta \mathbf{k} \mathbb{I} + \sum_{a=1,2,3} (\mathbf{v}_a . \delta \mathbf{k})~\sigma_a~, 
\end{equation}
where a three component vector $\mathbf{v}_{j} = \nabla_\mathbf{k} f_{j}(\mathbf{k}) |_{\mathbf{k}= \pm \mathbf{k}^{-}_{\rm W}}$ with $j = 0,1,2,3$. The chirality of each Weyl node is given by $C = sign(\mathbf{v}_x. \mathbf{v}_y \times \mathbf{v}_z)$ \cite{Ashvin2018}. For CaAuAs, we find the chirality of Weyl points located at $(+ k_{\rm W}^{+}, + k_{\rm W}^{-}, - k_{\rm W}^{-}, - k_{\rm W}^{+} ) $ is $(+1,-1,+1,-1)$ which is shown in Fig.~\ref{fig:WS}(d). Consequently, the Fermi arcs associated with each Dirac node naturally provide the Fermi arcs connecting a pair of Weyl nodes as illustrated schematically in Fig.~\ref{fig:WS}(d). 

\section{Conclusion} \label{Conc}
Based on our first-principles band structure calculations and an effective low-energy model Hamiltonian analysis, we predict that, without the SOC, CaAuAs hosts a unique starfruit-like nodal link semimetal state, which is formed by three equivalent nodal loops which link together at the high-symmetry point on the $k_z$ axis. When the SOC is tuned on, the nodal lines are gapped and CaAuAs realizes a $C_3$ rotation-symmetry-protected Dirac semimetal with a pair of Dirac cones. These Dirac cones exhibit a linear dispersion along all momentum directions. The surface spectrum reveals the existence of a pair of double Fermi arc states, which connect the projected Dirac cones over the surface. We also discuss how a variety of topological states emerge from the Dirac semimetal when various symmetries are broken. In particular, we show that the breaking of $C_3$ rotational symmetry drives CaAuAs into a topological insulating state, while breaking of the time-reversal symmetry leads to a Weyl semimetal with two pairs of Weyl points. We thus conclude that CaAuAs offers an interesting platform for investigating the exotic quantum phenomena.

\section*{ACKNOWLEDGMENTS}
Work at the ShenZhen university is financially supported by the Shenzhen Peacock Plan (Grant No. 827-000113, KQTD2016053112042971), Science and Technology Planning Project of Guangdong Province (2016B050501005), and the Educational Commission of Guangdong Province (2016KSTCX126). The work at Northeastern University was supported by the US Department of Energy (DOE), Office of Science, Basic Energy Sciences grant number DE-FG02-07ER46352, and benefited from Northeastern University's Advanced Scientific Computation Center and the National Energy Research Scientific Computing Center through DOE grant number DE-AC02-05CH11231.

\bibliographystyle{prsty}
\bibliography{CaAuAs}
\end{document}